\documentclass[]{gji}
\usepackage{timet}
\usepackage[dvipdfmx]{graphicx}

\title[Atomistic derivation of rate and state friction law]
{Rate and state friction law as derived from atomistic processes at asperities}
\author[T. Hatano]
  {Takahiro Hatano$^1$\\
  $^1$ Earthquake Research Institute, The University of Tokyo,
    Yayoi, Bunkyo \emph{1130032} Tokyo, Japan}
\date{}
\pagerange{\pageref{firstpage}--\pageref{lastpage}}
\volume{}
\pubyear{}

\begin{document}
\label{firstpage}
\maketitle

\begin{summary}
A theoretical account is given of the microscopic basis of the rate- and state-dependent friction (RSF) law.
The RSF law describes rock friction quantitatively and therefore it is commonly used to model earthquakes and the related phenomena.
But the RSF law is rather empirical and the theoretical basis has not been very clear.
Here we derive the RSF law starting from constitutive laws for asperities, and give the atomistic expressions for the empirical RSF parameters.
In particular, we show that both the length constant and the state variable are given as the {\it 0th weighted power means} of the corresponding microscopic quantities: a linear dimension and the contact duration of each asperity.
As a result, evolution laws for the state variable can be derived systematically.
We demonstrate that the aging and the slip laws can be derived and clarify the approximations behind these two major evolution laws.
Additionally, the scaling properties of the length constant are clarified for fractal distribution of asperities.
\end{summary}


\section{Introduction}
\subsection{Rate- and State-Dependent Friction Law}
Rock friction may be regarded as an elementary ingredient of the crustal deformation processes including earthquakes and therefore it has been extensively investigated mainly by means of laboratory experiment.
The essential progress has been brought by \cite{Dieterich1979}, who clarified the nature of frictional force for both steady states and transient states and formulated an empirical friction law.
\cite{Ruina1983} modified the original form to its present form, which is now referred to as the rate- and state-dependent friction (RSF) law.

In the framework of the RSF law, the friction coefficient depends on the slip velocity and some time-dependent variables that describe the physical state of the rubbing surfaces.
These time-dependent variables are referred to as the {\it state variables}.
In the simplest case, the friction coefficient $\mu$ is described as a function of the slip velocity $V$ and a single state variable $\theta$.
\begin{equation}
\label{rsf}
\mu (V, \theta) = \mu_* + a\log\frac{V}{V_*} + b\log\frac{\theta}{\theta_*},
\end{equation}
where $\mu_*$ is the steady-state friction coefficient at a reference slip velocity $V_*$; $\theta_*$ the steady-state state variable at $V_*$; and $a$ and $b$ positive nondimensional constants.
In typical experiments, $a$ and $b$ are on the order of $0.01$ \cite{Ruina1983,Tullis1986,Linker1992}.

On the right hand side of Eq. (\ref{rsf}), the second term describes the velocity dependence of friction, and the third term describes the state-variable dependence.
The state variable $\theta$ are responsible for all the ingredients that affect friction other than the slip velocity $V$.
For instance, the time-dependent behaviour of friction must be described through the state variable, and therefore a time evolution law for the state variable is needed in addition to Eq. (\ref{rsf}).
Among many empirical evolution laws that have been proposed so far, the following two equations are most commonly used \cite{Ruina1983}.
\begin{eqnarray}
\label{dieterich}
\dot{\theta} &=& 1 - \frac{V}{L}\theta,\\
\label{ruina}
\dot{\theta} &=& \frac{V\theta}{L} \log\frac{L}{V\theta},
\end{eqnarray}
where $L$ is the length constant.
These equations are referred to as the aging law and the slip law, respectively.
Although there have been many attempts to settle the most suitable evolution law \cite{Beeler1994,Perrin1995,Kato2001}, no decisive conclusions have been made \cite{Marone1998}.
One should also pay attention to a recent attempt to devise a novel evolution law \cite{Nagata2012}.

Irrespective of the choice of evolution law, the steady-state state variable at slip velocity of $V$ is given as $L/V$, and therefore the steady-state friction coefficient is given by 
\begin{equation}
\label{steadystate}
\mu(V, \theta_{\rm ss}) \equiv \mu_{\rm ss} (V)= \mu_* + (a-b) \log\frac{V}{V_*}.
\end{equation}
The coefficient for the logarithmic dependence, $a-b$, may be either positive or negative depending on the experimental conditions and the rock species \cite{Stesky1974,Marone1990,Blanpied1991,Reinen1994,Blanpied1998}.
Particularly, the sign of $a-b$ is closely related to the stability of steady-sliding state \cite{Ruina1983,Rice1983} and therefore plays a vital role in the literature of earthquake dynamics.

On the other hand, many experiments indicate that the RSF law is not valid if the slip velocity is sufficiently high ($V\ge 1$ mm/s) under high normal stress ($\ge 1$ MPa) \cite{Tsutsumi1997,diToro2011,Han2011}.
Therefore, application of the RSF law in its present form should be limited to quasistatic-to-intermediate slip velocities.
Nevertheless, the RSF law is still important in describing various phenomena in which the slip is quasistatic: the nucleation process of earthquake \cite{Dieterich1992,Dieterich1994a,Ampuero2008}, afterslip \cite{Marone1991}, and slow slip events \cite{Liu2009,Rubin2008,Hawthorne2013}.

Interestingly, Eq. (\ref{rsf}) together with an appropriate evolution law describes the behavior of friction coefficient not only for rocks but also for paper \cite{Heslot1994}, steel \cite{Popov2012}, wood, glass, and acrylic plastic \cite{Dieterich1994}.
In this sense, the framework of the RSF law is universal and therefore may be important to much wider literatures.
Actually, the logarithmic dependences in Eq. (\ref{rsf}) can be derived without any prior knowledge on the underlying physical processes \cite{Hatano2015}.

\subsection{Physical meaning of parameters}
The RSF and evolution laws, Eqs. (\ref{rsf}), (\ref{dieterich}), and (\ref{ruina}), involve three important parameters: $a$, $b$, and $L$.
The values of $a$ and $b$ (especially the sign of $a-b$) determine the stability of steady-sliding states \cite{Ruina1983,Rice1983}.
The length constant $L$ may be regarded as the characteristic slip length, over which the friction force relaxes to the steady-state value.
Typical estimate of $L$ in laboratory experiments is on the order of micrometers \cite{Ruina1983,Tullis1986,Linker1992}.
In the literature of earthquake generation, the length constant may scale the size of critical nucleation size of rupture, above which the crack growth becomes unstable \cite{Dieterich1992}.
Therefore, if one wishes to understand the earthquake dynamics based on the RSF law, determination of the parameter values is crucial.

The nondimensional parameters, $a$ and $b$, are relatively well understood from the atomistic point of view.
For instance, by assuming a thermal activation process for sliding, one can easily obtain $a=k_BT/P\Omega$ \cite{Chester1994,Heslot1994,Baumberger1997,Baumberger1999,Rice2001}, where $k_B$ is the Boltzmann constant, $T$ the temperature, $P$ the normal stress on the real contact area, and $\Omega$ the activation volume.
This expression seems to be verified experimentally \cite{Nakatani2001}.
In a similar manner, an atomistic expression for $b$ is obtained by \cite{Bar-Sinai2014}.

On the other hand, the length constant $L$ is rather poorly understood.
It is typically on the order of $1 \mu$m in the laboratory-scale.
But the parameter estimated by geodetic inversion assuming the RSF law for plate boundaries is on the order of milimeters \cite{Fukuda2009} to centimeters \cite{Kano2015}.
There is a discrepancy of more than three orders of magnitude between laboratory and observation.
This may indicate the scale-dependence of the RSF law, but we do not have any clear explanation on this wide discrepancy.
Because the RSF law is an empirical law obtained by laboratory experiments only and lacks a theoretical derivation, one cannot deduce a parameter value of $L$ beyond the laboratory scale.

The contact between two macroscopic surfaces is accommodated by a set of microscopic junctions \cite{Bowden}, an average dimension of which is typically $\sim 1 \ \mu$m \cite{Dieterich1996}.
Throughout this paper, we refer to such junctions as {\it asperities}.
The conventional interpretation of the length constant $L$ is a ''typical'' dimension of the asperities \cite{Yoshioka1996}.
However, the definition of ''typical'' is not clear if the size of asperity is distributed over a wide range.
Actually, the size distribution of asperity is a power law for rough brittle surfaces \cite{Dieterich1996}.

Let us assume that the asperity size distribution $\rho(l)$ is proportional to $l^{-\alpha}$ with the lower and the upper cutoffs being $L_{\rm min}$ and $L_{\rm max}$, respectively.
These cutoff length constants may be candidates of the typical length, but one cannot choose one and discard the other with legitimate reasoning because there has not been a mathematical expression for the typical length.
For instance, if we were to adopt the arithmetic average as the typical length, $L$ would be proportional to $L_{\rm min}$ for $\alpha>2$.
Actually, a direct observation of asperities revealed that the exponent $\alpha$ can be larger than $2$ \cite{Dieterich1996}.
In this case $L$ would be the smaller length scale $L_{\rm min}$ irrespective of the upper cutoff $L_{\rm max}$.
Because $L_{\rm min}$ may be determined by the local wear process on the frictional surface rather than global faulting dynamics, we could expect that $L_{\rm min}$ is a microscopic length scale that is independent of the system size.
Therefore, if $L$ is defined by the arithmetic average, $L$ would remain a microscopic length scale and there would not be scale-dependence.
To resolve this paradoxical situation, we need a clear mathematical expression for $L$.


\subsection{Aim of This Paper}
To clarify the mathematical definition of $L$ and to resolve the scale-dependent nature, one must derive the RSF law theoretically and establish the connection between the empirical friction law and the microscopic physical processes.
In doing so, one can clarify the physical origin and the scaling properties of the parameters and provide them with the theoretical basis.

In section \ref{section:single}, the atomistic nature of a single asperity is discussed and the constitutive laws are proposed.
Based on these discussions, in section \ref{section:derivation}, the RSF law is derived and the microscopic expressions for the empirical parameters are given.
Importantly, the microscopic expression for the state variable is also clarified.
In section \ref{section:parameters}, the quantitative details and the statistical nature of the parameters are discussed.
In particular, quantitative constraints on some parameters are given under the condition that the RSF law is valid.
In section \ref{section:WPM}, the scaling properties of the length constant $L$ are discussed.
In section \ref{section:evolutionlaw}, evolution laws are derived systematically based on the microscopic expression for the state variable.
Here we limit ourselves to two major laws: the aging and the slip laws.
Through the derivation, one can understand the nature of approximations behind each evolution law.
In section \ref{section:alternative}, we formulate an alternative form of friction law.
In contrast to the conventional RSF law, the formulation given here does not refer to any reference state and therefore we can determine the absolute value of the friction coefficient in terms of the material constants.
Section \ref{section:conclusions} is devoted to concluding remarks.

\section{Atomistic processes and constitutive laws at asperity}
\label{section:single}
Throughout this paper, we adopt the natural logarithm for mathemetical convenience.
Then, $a$ and $b$ here must be multiplied by $\log10$ to be comparable with those for the common logarithm, which is widely used in the literature.
The length constant $L$ does not depend on the base of the logarithm.

\subsection{General remarks}
An essential microscopic ingredient of friction is asperity, which is a nominal junction of protrusions of the surfaces \cite{Bowden,Rabinowicz1965}.
Throughout this paper, we write the area of and the shear stress at asperity $i$ as $A_i$ and $\sigma_i$, respectively.
Then the macroscopic friction force $F$, which is just the sum of traction on the asperities, may be written as
\begin{equation}
\label{start}
F = \sum_{i\in{\cal S}} \sigma_i A_i,
\end{equation}
where $\cal S$ denotes a set of asperities for given contacting surfaces.
Note that Eq. (\ref{start}) is an identity and must be valid even if the slip is spatially nonuniform over the sliding interface.
The total area of asperities defines the real contact area, which is denoted by $A_{\rm real}$.
\begin{equation}
\label{Areal}
A_{\rm real} \equiv \sum_{i\in{\cal S}} A_i.
\end{equation}
Generally, $A_{\rm real}$ is much smaller than the apparent contact area \cite{Bowden,Dieterich1994,Pei2005}.

Equation (\ref{start}) means that the nature of macroscopic friction force is determined by the rheological properties of asperity, $\sigma_i$.
This involves the atomistic processes that occur at sliding asperities, and therefore one must pay due attention to the atomistic structure of sliding asperities.
If two protrusions that constitute an asperity retain their crystalline structure, the asperity may be regarded as a grain boundary.
If such asperities are common, friction may be regarded as grain-boundary sliding from a microscopic point of view.
Although the atomistic structure of a grain boundary and the nature of grain-boundary sliding may depend on the extent of lattice misorientation at the interface, it is only true for low-angle grain boundaries and some special high-angle grain boundaries.
In most cases we could expect that the rheological properties of grain-boundary sliding is insensitive to the nature of misorientation.

If an asperity looses its original crystalline nature and is regarded as amorphous, the rheological properties of a single asperity may be isotropic as well.
Therefore, anisotropy due to the crystalline structure of asperities may be negligible especially if we consider the averaged properties for many asperities.


\subsection{Constitutive law for asperity}
Irrespective of the atomistic nature of asperity, we assume that breaking and reconnection of the covalent bonds determine the shear stress at asperities.
Throughout this paper, we assume that the covalent-bond reconnection is a thermal activation process and the rate of reconnection is described by the Arrhenius law.
Additionally, we assume that the local shear rate at asperity is proportional to the bond-reconnection rate.
The proportional coefficient is denoted by $\gamma_0$, which is regarded as the average strain needed for bond reconnection.
Then we can write the following relation between the shear rate $\dot\gamma$ and the amplitude of interatomic force $f_{\alpha}$ acting on a single covalent bond $\alpha$.
\begin{equation}
\label{sinh}
\dot\gamma = 2\omega_D \gamma_0\exp\left(-\frac{\epsilon_{\alpha}}{k_BT}\right) \sinh\left(\frac{f_{\alpha}l}{k_BT}\right),
\end{equation}
where $\omega_D$ is the attempt frequency, $k_B$ the Boltzmann constant, $T$ the temperature, $l$ a length constant, $\epsilon_{\alpha}$ the activation energy for the reconnection process of covalent bond $\alpha$.
Note that the activation energy may depend on $\alpha$ due to the elastic interaction between bond $\alpha$ and the neighboring sites \cite{Liu2012}.
Throughout this paper, we assume that $f_{\alpha}l \ge k_BT$.
Then Eq. (\ref{sinh}) is approximated as
\begin{equation}
\label{exponential}
\dot\gamma = \gamma_0\omega_D\exp \left(\frac{f_{\alpha}l-\epsilon_{\alpha}}{k_BT}\right).
\end{equation}
The idea that friction may be a thermal activation process has been pointed out by \cite{Chester1992,Chester1994} and \cite{Heslot1994}.

Next we consider an asperity sliding at the velocity of $V$.
Using the shear-zone thickness at asperities, $d$, the shear rate $\dot\gamma$ is written as $V/d$.
Then  from Eq. (\ref{exponential}) one can immediately obtain
\begin{eqnarray}
\label{f_alpha}
f_{\alpha} &=& \frac{\epsilon_{\alpha}}{l} + \frac{k_BT}{l}\log\left(\frac{V}{V_0}\right),\\
\label{V_0}
V_0&\equiv& \gamma_0  \omega_D d.
\end{eqnarray}
The shear stress at asperity $i$ reads
\begin{equation}
\label{sigma_i}
\sigma_i = \sum_{\rm unit} f_{\alpha} \simeq n_i {\bar f},
\end{equation}
where the summation is taken over the covalent bonds per unit area of asperity $i$, $n_i$ the areal density of the covalent bonds, and ${\bar f}$ the average force on a single covalent bond.
One gets from Eq. (\ref{f_alpha})
\begin{equation}
\label{activation}
{\bar f} \equiv \frac{E}{l} + \frac{k_BT}{l}\log\left(\frac{V}{V_0}\right),
\end{equation}
where $E$ is the average of $\epsilon_{\alpha}$.

Equation (\ref{activation}) contains several atomistic parameters, but we do not discuss the quantitative details for specific materials and limit ourselves to an order-of-magnitude estimate.
\begin{enumerate}
\item It follows from the definition of thermal activation that $E \gg k_BT$.
Otherwise covalent-bond reconnection should not be regarded as a thermal activation process and therefore Eq. (\ref{sinh}) does not apply.
\item The length constant $l$ may be on the order of the interatomic distance because the energy scale ${\bar f} l$ is interpreted as the work done in a bond-reconnection event.
\item The attempt frequency of bond reconnection, $\omega_D$, may be the Debye frequency.
\item We assume that the shear-zone of asperities is atomistically thin: $d\simeq l$.
Then the velocity constant $\omega_D d$ may be interpreted as the sound velocity.
\item Average strain needed for a bond-reconnection event, $\gamma_0$, is set to be $10^{-2}$.
Then it follows from Eq. (\ref{V_0}) that $V_0$ is $1$ \% of the sound velocity.
\item Because $V_0$ is much larger than typical slip velocities, the second term on the right hand side of Eq. (\ref{activation}) is negative and therefore $E > {\bar f} l$.
\item Because $E \gg k_BT$, we have the following inequalities for the parameters.
\end{enumerate}
\begin{equation}
\label{inequality_general}
k_BT \le {\bar f} l < E.
\end{equation}

\subsection{Aging of asperity}
The essential ingredient of friction is aging: This is responsible not only for time-dependent increase of static friction but also for the negative velocity dependence of dynamic friction.
The current popular belief on the microscopic physical process behind frictional aging is the time-dependent increase of the real contact area, which has been directly observed using transparent materials \cite{Dieterich1994,Dieterich1996}.
This may be modeled by introducing the time dependence in $A_i$.
However, an experiment on nanoscale contact reveals that frictional aging can occur even if the real contact area is constant \cite{Li2011}.
This may be attributed to the increase of covalent bond density at the interface as demonstrated by simulations \cite{Liu2012}.
This may be modeled by introducing the time dependence in $n_i$.
One could also consider the case in which both processes occur at the same time: the bond density and the real contact area increase with time.
In any case, however, one can model the frictional aging in a unified manner using the total number of the covalent bonds at asperity $i$, which is denoted by $Z_i$.
\begin{equation}
\label{define_Z}
Z_i \equiv n_i A_i.
\end{equation}
We then adopt the following time dependence in $Z_i$.
\begin{equation}
\label{loghealing}
Z_{i} (\theta_i) = Z_i(0) \left[1+c\log\left(1+\frac{\theta_i}{\tau}\right)\right],
\end{equation}
where $\theta_i$ is the the duration of contact of asperity $i$, $\tau$ the characteristic time for healing, and $c$ the nondimensional coefficient.
Note that Eq. (\ref{loghealing}) may be valid irrespective of the physical processes for aging.

The parameters $c$ and $\tau$ in Eq. (\ref{loghealing}) play a central role in the RSF law through frictional aging:
The characteristic time $\tau$ limits the aging for shorter time scale, and the nondimensional constant $c$ determines the extent of aging.
The experiment on amorphous silica using an atomic force microscope shows that $\tau$ is approximately $0.1$ s \cite{Li2011}.
A phenomenological constant $c$ has been measured by two independent methods: the waiting-time dependence of static friction and direct observation of the real contact area.
While the former method gives $c\sim0.01$ \cite{Baumberger1997,Baumberger1999}, it could be somewhat problematic because the static friction may depend on the spatiotemporal dynamics of the slip initiation \cite{Ben-David2011}.
Therefore, static friction itself may be scale- and geometry-dependent.
Actually, an experiment on a single nano-asperity using an atomic force microscope shows $c\sim 1$.
However, because the latter method also gave $c\sim0.01$ \cite{Dieterich1996}, the actual value for $c$ may not be so decisive.

Although the parameters $c$ and $\tau$ in Eq. (\ref{loghealing}) themselves should be expressed using the material constants, for this purpose one must identify the atomistic mechanism of aging.
For instance, in the case of plastic creep of asperity under uniaxial compression, \cite{Brechet1994} obtains 
\begin{eqnarray}
A_{i} (\theta_i) &=& A_i(0)\left[1+c\log\left(1+\frac{\theta_i}{\tau}\right)\right],\\
\label{c}
c &=& \frac{k_BT}{P\Omega'},\\
\label{tau}
\tau &=& f_D^{-1} \frac{k_BT}{P\Omega'}\exp\left(\frac{E' - P \Omega'}{k_BT} \right),
\end{eqnarray}
where $P$ is the normal stress at the asperities, $f_D$ the Debye frequency, and $E'$ and $\Omega'$ the activation energy and the activation volume for plastic creep, respectively.
The activation energy $E'$ is not necessarily the same as that for covalent bond reconnection because the atomistic mechanisms may be different in general.
The normal stress $P$ may be the yield stress under uniaxial compression.
Provided that $P=8$ GPa and $\Omega'=4\times10^{-29}$ ${\rm m}^3$, one can estimate that the constant $c$ is approximately $10^{-2}$.
On the other hand, quantitative estimate of $\tau$ is not straightforward.
This is partially because $\tau$ is very sensitive to the estimate of activation energy $E'$ as is apparent in Eq. (\ref{tau}).


\section{Derivation of RSF law} \label{section:derivation}
In this section we consider macroscopic interfaces that possess a large number of asperities.
In general, the friction of such macroscopic interfaces may not be uniform; namely, the slip velocity can be spatially inhomogeneous.
Even in such cases, however, we may expect that there is a mesoscopic length scale in which the slip velocity may be approximated as uniform.
We focus on such a mesoscopic region and write 
\begin{equation}
\label{fZ}
F={\bar f} \sum_{i\in{\cal S}} n_i A_i = {\bar f} \sum_{i\in{\cal S}} Z_i,
\end{equation}
where we use Eqs. (\ref{start}), (\ref{sigma_i}), and (\ref{activation}).
Note that the summation here is taken for a region where the slip velocity is regarded as uniform.
From Eq. (\ref{fZ}) we derive the RSF law based on the microscopic properties discussed in the previous section.

Because the RSF law is cast as the difference from a reference state of $(V_*, \theta_*)$, the derivation must also involve a reference state.
We take the steady state at slip velocity of $V_*$ as the reference state and write ${\bar f}$ and $Z_i$ in the form of deviation from the steady state: ${\bar f} ={\bar f} (V_*) +\Delta{\bar f}$ and $Z_i = Z_i^{\rm ss} (V_*) + \Delta Z_i$.
Here the superscript "ss" indicates the steady-state value.
We expect that the relaxation time for ${\bar f}$ is negligible and therefore ${\bar f}$ always takes the steady-state value described by Eq. (\ref{activation}).
Then the macroscopic friction force is described as 
\begin{equation}
\label{rsf_formal}
F \simeq F_* + \Delta{\bar f} \sum_{i\in{\cal S}} Z_i^{\rm ss} (V_*)
 + {\bar f}(V_*) \sum_{i\in{\cal S}}\Delta Z_i,
\end{equation}
where $F_* =  {\bar f}^{\rm ss} (V_*)\sum_{i} Z_i^{\rm ss} (V_*) $.
Note that the second order term is neglected; and for this purpose a reference velocity $V_*$ must be a typical value used in experiments.
This condition is important in the interpretation of the parameters.

Then we calculate each term in Eq. (\ref{rsf_formal}). Obviously, using Eq. (\ref{activation}),
\begin{equation}
\label{delta_f}
\Delta{\bar f} = {\bar f}(V)-{\bar f}(V_*) = \frac{k_BT}{l} \log\left(\frac{V}{V_*}\right).
\end{equation}

From  Eq. (\ref{loghealing}) one can compute
\begin{equation}
\label{Z_1}
\sum_{i\in{\cal S}} Z_{i}(\theta_i)= \left[\sum_{i\in{\cal S}} Z_i(0)\right]
\left\{ 1 + c \log\left[\prod_{i\in {\cal S}}\left(1+\frac{\theta_i}{\tau}\right)^{\xi_i}\right]\right\},
\end{equation}
where
\begin{equation}
\label{xi_Z}
\xi_i \equiv \frac{Z_i(0)}{\sum_i Z_i(0)}.
\end{equation}
Noting that $\sum\xi_i=1$, one can further rewrite Eq. (\ref{Z_1}) as
\begin{equation}
\label{Z}
\sum_{i\in{\cal S}} Z_{i}(\theta_i) = \left[\sum_{i\in {\cal S}} Z_i(0)\right] \left(1 + c\log\frac{\theta}{\tau}\right),
\end{equation}
where
\begin{equation}
\label{theta}
\theta \equiv \prod_{i\in{\cal S}} (\theta_i + \tau) ^{\xi_i}.
\end{equation}

Similarly, one can calculate $\sum Z_i^{\rm ss}(V_*)$.
In a steady-sliding state, the contact duration of an asperity may be written as 
\begin{equation}
\label{ti_ss}
\theta_i^{\rm ss} (V_*) \simeq \frac{L_i}{V_*},
\end{equation}
where $L_i$ is a linear dimension of asperity $i$ along the slip direction.
Inserting Eq. (\ref{ti_ss}) into Eq. (\ref{Z_1}) and following the same manner as one obtains Eq. (\ref{Z}), one is led to the total bond number for the steady state at slip velocity of $V_*$.
\begin{equation}
\label{Zss}
\sum_{i\in{\cal S}} Z_i^{\rm ss}(V_*)= \left[\sum_{i\in{\cal S}} Z_i (0) \right] \left(1+c\log\frac{L}{V_*\tau}\right),
\end{equation}
where
\begin{equation}
\label{L}
L \equiv \prod_{i\in{\cal S}} (L_i + V_* \tau)^{\xi_i}.
\end{equation}

Subtracting Eqs. (\ref{Zss}) from (\ref{Z}),  one obtains
\begin{equation}
\label{deltaZ}
\sum_{i\in{\cal S}} \Delta Z_i \simeq c \left[\sum_{i\in{\cal S}} Z_i (0) \right]\log \frac{V_*\theta}{L}.
\end{equation}
Note that the asperity sets ${\cal S}$ in Eqs. (\ref{Z}) and (\ref{Zss}) are not strictly the same because the detailed configuration of asperities may change due to sliding.
For simplicity, however, we neglect the difference in the asperity sets in obtaining Eq. (\ref{deltaZ}).

Then, to derive the RSF law, one just needs to insert Eqs. (\ref{activation}), (\ref{delta_f}), (\ref{Zss}) and (\ref{deltaZ}) into Eq. (\ref{rsf_formal}), and divide it by the normal load $N$.
\begin{equation}
\label{rsf2}
\mu \simeq \mu_*+a \log\frac{V}{V_*} + b \log\frac{V_*\theta}{L}, 
\end{equation}
where
\begin{eqnarray}
\label{micro_a}
a &=& \frac{k_BT}{Nl}\left[\sum_{i\in{\cal S}} Z_i (0)\right] \left(1+c\log\frac{L}{V_*\tau}\right),\\
\label{micro_b}
b &=& c \frac{k_BT}{Nl}\left[\sum_{i\in{\cal S}} Z_i (0) \right]
\left(\frac{E}{k_BT} + \log\frac{V_*}{V_0}\right).
\end{eqnarray}
Equation (\ref{rsf2}) is equivalent to Eq. (\ref{rsf}), but now we have microscopic expressions for the parameters: Eqs. (\ref{theta}), (\ref{L}), (\ref{micro_a}), and (\ref{micro_b}).

Note that Eqs. (\ref{theta}) and (\ref{L}) give microscopic expressions for the state variable $\theta$ and the length constant $L$, respectively.
From these equations together with Eq. (\ref{ti_ss}), one immediately retains the state variable at a steady state.
\begin{equation}
\label{theta_ss}
\theta_{\rm ss} = V^{-1} \prod_{i\in{\cal S}} (L_i + V \tau)^{\xi_i} = \frac{L}{V}.
\end{equation}
By inserting Eq. (\ref{theta_ss}) into (\ref{rsf2}), one retains the steady-state RSF law, Eq. (\ref{steadystate}).

\section{Theoretical expressions for parameters}\label{section:parameters}
In deriving the RSF law, we obtain the microscopic expressions for the state variable $\theta$ as Eq. (\ref{theta}) and the length constant $L$ as Eq. (\ref{L}).
The nondimensional parameters $a$ and $b$ are given by Eqs. (\ref{micro_a}) and (\ref{micro_b}).
In this section, some variations of these expressions are presented together with their physical meaning.

\subsection{Nondimensional parameters $a$ and $b$}
One can get simpler expressions for $a$ and $b$ with a further assumption that $n_i(0)=n_0$:
The areal density of covalent bonds depends only on the contact duration $\theta_i$.
With this assumption, one has
\begin{equation}
\label{n_0}
Z_i(0) \simeq n_0 A_i(0),
\end{equation}
where we use Eq. (\ref{define_Z}).
Then Eqs. (\ref{micro_a}) and (\ref{micro_b}) become
\begin{eqnarray}
\label{a}
a &=& \frac{k_BT}{P\Omega} \left(1+c\log\frac{L}{V_*\tau}\right),\\
\label{b}
b &=& \frac{cE}{P\Omega}\left(1 + \frac{k_BT}{E}\log\frac{V_*}{V_0}\right).
\end{eqnarray}
where 
\begin{eqnarray}
\label{P}
P &\equiv& \frac{N}{\sum_i A_i(0)},\\
\label{Omega}
\Omega &\equiv& \frac{l}{n_0}.
\end{eqnarray}
Here $P$ is interpreted as the average normal stress at asperities, which is approximately the indentation strength of a material.
We refer to it as the {\it real normal stress}.
The latter, $\Omega$, has the dimension of volume, and we refer to it as the {\it activation volume}.

One can get even simpler expression for $b$.
Noting that $V_*$ must be a typical slip velocity used in experiments, one can estimate from Eq. (\ref{activation}) that
\begin{equation}
V_* \sim V_0 \exp\left(\frac{{\bar f}l - E}{k_BT}\right).
\end{equation}
Inserting the above equation to Eq. (\ref{micro_b}) and noting that $\mu=F/N={\bar f}\sum_i Z_i/N$, one obtains
\begin{equation}
\label{b_simple}
b \simeq c \mu_*,
\end{equation}
where $\mu_*$ is the steady-state friction coefficient at slip velocity of $V_*$.
This ensures that $b\sim c$ because $\mu_*\sim 0.7$ for typical rocks and minerals.

As apparent in Eqs. (\ref{a}) and (\ref{b}), the parameters $a$ and $b$ could depend on a reference velocity $V_*$.
Then they would not be objective because $V_*$ may be chosen arbitrarily within a range of slip velocities used in experiment.
But we can show that their dependence on $V_*$ may be negligible.

First, one has to recall that $V_*$ must be a typical experimental value for Eq. (\ref{rsf_formal}) to be valid.
Noting that $L$ is on the order of micrometers, $L/V_*$ may be on the order of $10^{-3}$ to $10^3$ seconds.
The characteristic time for frictional aging, $\tau$, is estimated as approximately $0.1$ seconds \cite{Li2011}.
For convenience, the values used here are tabulated in Table I.
Note that, however, some of the estimates are crude and lack the experimental validation.
For instance, we assume $c\sim 0.01$ based on the literature of rock friction experiments \cite{Dieterich1994,Baumberger1999}, whereas it could be on the order of $1$ for an nanoasperity realized in  an atomic force microscopy \cite{Li2011}.
However, if we adopted $c\sim 1$, the RSF law would hardly hold as we show below.
Additionally, there is orders-of-magnitude uncertainty in the velocity constant $V_0$, but this may not cause a serious problem because $V_0$ appears only in the form of logarithm.

\begin{table}[ht]
\begin{center}
\caption{Numerical values for constants used for the estimation of the RSF parameters.
Note that $c$ must be multiplied by $\log10$ if one adopts the common logarithm.}
  \begin{tabular}{llr}\hline \hline
    $c$ & (aging constant) & 0.01  \\
  $L$ & (length constant) & $10^{-6}$ m \\
    $\tau$ & (time constant for aging) & $0.1$ s \\
      $V_0$ & (velocity constant) &  $30$ m/s \\
  $V_*$ & (reference slip velocity) & $10^{-6}$ m/s \\
  \hline\hline
  \end{tabular}
\end{center}
\end{table}

Because $c$ is approximately $0.01$ and $\log(L/V_*\tau)$ is on the order of $1$, the second term in the bracket on the right hand side of Eq. (\ref{a}) is negligible.
Thus, one gets
\begin{equation}
\label{a_simple}
a \simeq \frac{k_BT}{P\Omega},
\end{equation}
which is consistent with the previous studies \cite{Baumberger1999,Nakatani2001,Rice2001}.
If $c$ were on the order of $1$, the second term on the right hand side of Eq. (\ref{a}) could not be negligible and  parameter $a$ would depend on an arbitrary reference velocity $V_*$.
In this respect, $c$ must be much smaller than $1$.

On the other hand, parameter $b$ is not so straightforward.
The second term in the bracket on the right hand side of Eq. (\ref{b}) would be negligible if $E/k_BT \ge 170$, whereas $\log(V_*/V_0)\simeq -17$.
If so, Eq. (\ref{b}) becomes
\begin{equation}
b \simeq c\frac{E}{P\Omega},
\end{equation}
as is already given by \cite{Bar-Sinai2014}.
However, such a large activation energy is not plausible:
In some previous studies, the activation energy $E$ has been estimated as $180$ kJ/mol \cite{Nakatani2001,Rice2001}, which gives $E/k_BT=70$ even at $T=300$ K.
Therefore, the $\log(V_*/V_0)$ term in Eq. (\ref{b}) itself may not be negligible.
This means that $b$ cannot be expressed in terms of material constants only.

In a practical respect, however, this does not cause a serious problem because the change of one decade in $V_*$ yields only $10$ \% change in $b$.
Such a small change could be detected in a precise measurement, but so far we are unaware of such experiments.
To show this, let us write $b$ as $b(V_*)$ to show the velocity dependence explicitly and define $\Delta b\equiv b(V_*')-b(V_*)$.
This is the relative change of $b$, which is written as
\begin{equation}
\label{relative_db}
\frac{\Delta b}{b}=\frac{\log\frac{V_*'}{V_*}}{\frac{E}{k_BT}+\log\frac{V_*}{V_0}}.
\end{equation}
Inserting the values shown in Table I and $E/k_BT=70$ into the above equation, one can confirm that $\Delta b/b$ is less than $0.1$ for $V_*'/V_*=10$.
Note that, however, the reference-state dependence of $b$ is more significant at higher temperatures.
For instance, at $T=600$ K (and $E/k_BT=35$), one-decade change in $V_*$ leads to $25$ \% change in $b$.
This may lead to more complex frictional properties than the standard RSF.

\subsection{Steady states}
From Eqs. (\ref{a}) and (\ref{b}), one obtains
\begin{equation}
\label{a-b}
a-b = \frac{k_B T}{P \Omega}\left(1-\frac{c E}{k_BT}+c\log\frac{LV_0}{V_*^2\tau}\right),
\end{equation}
which is equivalent to $\partial\mu_{\rm ss}/\partial(\log V)|_{V=V_*}$.
If the third term on the right hand side of Eq. (\ref{a-b}) was negligible, $a-b$ would be independent of the slip velocity $V_*$.
Using the values in table I, however, one can confirm that $\log(LV_0/V_*^2\tau)\sim 20$ and therefore it is rather comparable with $E/k_BT \simeq 70$.
On the other hand, the deviation from the logarithmic dependence on the slip velocity is not significant.
One can confirm this by differentiating Eq. (\ref{a-b}) with respect to $\log V$:
\begin{equation}
\frac{\partial (a-b)}{\partial \log V}= - 2 \frac{ck_BT}{P\Omega}.
\end{equation}
Because $c\ll 1$ and $k_BT \ll P\Omega$ as discussed in section \ref{activation_quantitative}, i.e., Eq. (\ref{kTPOmega}), the right hand side may be negligible .

The condition for velocity strengthening, namely, $a-b>0$, reads
\begin{equation}
\frac{E}{ k_BT} < \frac{1}{c} +\log\frac{LV_0}{V_*^2\tau}.
\end{equation}
From this equation one can see that the velocity strengthening is more common at higher temperatures.

\subsection{Constant $c$}
In the above discussions, the constant $c$ is assumed to be on the order of $0.01$.
If we assume $c\sim1$ as found for a nanoasperity \cite{Li2011}, the RSF law may not be valid for the following reasons:
(i) The parameter $a$ would depend on the reference velocity $V_*$ and cannot be regarded as a constant.
(ii) From Eq. (\ref{b_simple}), the constant $b$ would be on the order of $1$, which is too large to be comparable with any experiments.
Therefore, the assumption of $c\simeq0.01$ may be reasonable.


\subsection{Weight coefficient $\xi$}
The assumption of Eq. (\ref{n_0}) leads to an alternative expression of the weight coefficient $\xi_i$.
\begin{equation}
\label{xi}
\xi_i = \frac{A_i(0)}{\sum_i A_i(0)},
\end{equation}
which means that the weight coefficient is simply proportional to the area of asperity.
Because this is more intuitive than the other definition using the covalent bond density, Eq. (\ref{xi_Z}), hereafter we use Eq. (\ref{xi}) for the weight coefficient.

\subsection{State variable and length constant}
If the slip velocity is sufficiently low that $L_i \gg V_*\tau$ and $\theta_i\gg\tau$, Eqs. (\ref{theta}) and (\ref{L}) become
\begin{equation}
\label{theta_simple}
\theta \simeq \prod_{i\in{\cal S}} \theta_i ^{\xi_i},
\end{equation}
and
\begin{equation}
\label{L_simple}
L \simeq \prod_{i\in{\cal S}}L_i^{\xi_i}.
\end{equation}
Because the condition $\theta_i\gg\tau$ means that frictional healing is relevant, one can expect this condition applies as long as the RSF law is valid.
Equations (\ref{theta_simple}) and (\ref{L_simple}) reveal that the state variable $\theta$ and the length constant $L$ are {\it 0th weighted power mean} of their microscopic counterparts, $\theta_i$ and $L_i$.
The weight $\xi_i$ is given by Eq. (\ref{xi_Z}), or by Eq. (\ref{xi}) on the condition of Eq. (\ref{n_0}).

Mathematically, a 0th weighted power mean is smaller than or equal to a weighted mean \cite{Hardy}.
\begin{eqnarray}
\label{inequality_theta}
\theta \simeq \prod_{i\in{\cal S}} \theta_i^{\xi_i} &\le& \sum_{i\in{\cal S}}\xi_i \theta_i,\\
\label{inequality_L}
L \simeq \prod_{i\in{\cal S}} L_i^{\xi_i} &\le& \sum_{i\in{\cal S}}\xi_i L_i,
\end{eqnarray}
where an equality holds when and only when all the variables are equal; i.e., $\theta_i=t$ or $L_i=L$ for all $i$ in ${\cal S}$.
Thus, if the variance of $L_i$ (or $\theta_i$) is relatively small, the replacement of the 0th weighted power mean with the arithmetic average may not be a bad approximation.
However, as brittle surfaces may possess fractal topography in most experimental situations \cite{Dieterich1994}, such an approximation would require due verification.
This is discussed in section \ref{section:WPM}.

\section{Scale dependence of length constant $L$}\label{section:WPM}
As a microscopic expression for $L$ is obtained as Eqs. (\ref{L}) or (\ref{L_simple}), now we can clarify the scale dependence of $L$ focusing on the statistical properties of the 0th weighted power mean.
First, taking the logarithm of Eq. (\ref{L_simple}), one obtains
\begin{equation}
\label{logL_def}
\log L = \sum_i \xi_i \log L_i.
\end{equation}
Namely, the logarithm of $L$ is the weighted mean of $\log L_i$.
Recalling that the weight $\xi_i$ is the normalized asperity area given by Eq. (\ref{xi}), 
and that $A_i(0)$ may be proportional to the square of $L_i$, one obtains
\begin{eqnarray}
\label{logL}
\log L &\simeq& \sum_i  \tilde{\xi}_i\log L_i,\\
\tilde{\xi}_i &\equiv& \frac{L_i^2}{\sum_j L_j^2}.
\end{eqnarray}
Using the distribution function of asperity size $\rho(l)$, Eq. (\ref{logL}) can be rewritten as
\begin{equation}
\label{logL2}
\log L \simeq \frac{\int_{L_{\rm min}}^{L_{\rm max}} dl \rho(l) l^2 \log l}
{\int_{L_{\rm min}}^{L_{\rm max}} dl \rho(l) l^2}.
\end{equation}

\begin{figure}
\begin{center}
\label{fig:means}
\includegraphics[scale=0.5]{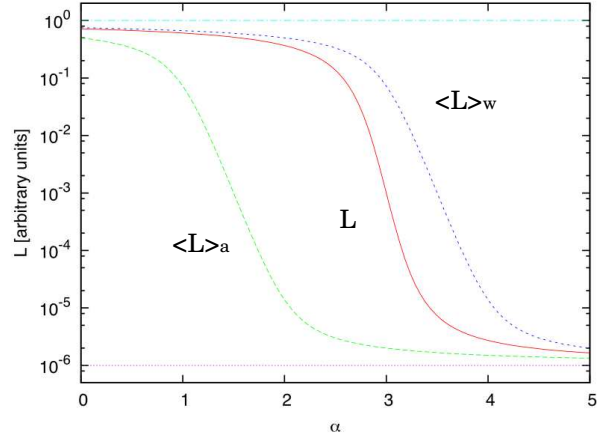}
\caption{
The 0th weighted power mean $L$ (the red solid line), the arithmetic mean $\langle L\rangle_{\rm a}$ (the green dashed line), and the weighted mean $\langle L\rangle_{\rm w}$ (the blue dotted line) of the asperity size.
The size distribution is assumed to be a power law; the exponent is $-\alpha$, $L_{\rm min}=10^{-6}$, and $L_{\rm max}=1$ (arbitrary units), respectively.
The two horizontal lines indicate $L_{\rm min}$ and $L_{\rm max}$, respectively.}
\end{center}
\end{figure}

If the distribution function of asperity size is power law, one inserts $\rho(l)\propto l^{-\alpha}$ into Eq. (\ref{logL2}) and obtains 
\begin{eqnarray}
\label{logL3}
\log L = \left\{ \begin{array}{ll}
\displaystyle \frac{L_{\rm max}^{3-\alpha}\log L_{\rm max}-L_{\rm min}^{3-\alpha}\log L_{\rm min}}
{L_{\rm max}^{3-\alpha}-L_{\rm min}^{3-\alpha}}-\frac{1}{3-\alpha} 
& \displaystyle (\alpha \neq 3),\\
\\
\displaystyle \frac{\log L_{\rm max}+\log L_{\rm max}}{2} & \displaystyle (\alpha = 3).\\
\end{array} \right.
\end{eqnarray}
Assuming that $L_{\rm max} \gg L_{\rm min}$, the above equations lead to
\begin{eqnarray}
\label{scalingL}
L \sim \left\{ \begin{array}{ll}
L_{\rm min} & (\alpha > 3),\\
\sqrt{L_{\rm min}L_{\rm max}} & (\alpha = 3),\\
L_{\rm max} & (\alpha  < 3).\\
\end{array} \right.
\end{eqnarray}
Obviously, $\alpha=3$ is the critical value for the length constant $L$.
If the exponent is larger than $3$, $L$ remains the microscopic length scale, $L_{\rm min}$.
The behavior of $L$ given by Eq. (\ref{logL3}) is shown in Fig. \ref{fig:means} as a function of $\alpha$.

In comparison, we also compute the arithmetic mean and the weighted mean, which are denoted by $\langle L\rangle_{\rm a}$ and $\langle L\rangle_{\rm w}$, respectively.
\begin{eqnarray}
\label{AM_L}
\langle L\rangle_{\rm a} &\equiv& \frac{\sum_i L_i}{\sum_j 1}
=\int_{L_{\rm min}}^{L_{\rm max}} dl \rho(l) l,\\
\label{WM_L}
\langle L\rangle_{\rm w} &\equiv& \sum_i \xi_i L_i
=\frac{\int_{L_{\rm min}}^{L_{\rm max}} dl \rho(l) l^3}{\int_{L_{\rm min}}^{L_{\rm max}} dl \rho(l) l^2},
\end{eqnarray}
where $\xi_i$ is given by Eq. (\ref{xi}).
These quantities are shown in Fig. \ref{fig:means} for a power law distribution $\rho(l)\propto l^{-\alpha}$.
Obviously, the arithmetic mean exhibits the significant deviation from the 0th weighted power mean.
If we were to adopt as $L$ the arithmetic mean of the asperity size, we would underestimate $L$ to a considerable amount for most cases; $0.5< \alpha <4$.
On the other hand, the weighted mean given by Eq. (\ref{WM_L}) may be a good approximation of the 0th weighted power mean except for $2.5< \alpha <4$.

Meanwhile, in experiments, one may also consider arbitrary distribution for asperity size.
For instance, we may consider bidisperse asperities: large asperities with a dimension of $L_1$ and small asperities with a dimension of $L_2$.
If the population ratio is $w:(1-w)$, $w_c=(L_2/L_1)^2$ is the critical ratio for the length constant $L$:
$L\simeq L_1$ for $w\gg w_c$, $L\simeq L_2$ for $w\ll w_c$, and $L=\sqrt{L_1L_2}$ for $w=w_c$.
Although such surfaces sould rather artificial, they might be more feasible in experiments to verify Eq. (\ref{L_simple}) than the systematic control of $\alpha$ for fractal surfaces.

\section{Derivation of Evolution Laws}\label{section:evolutionlaw}
Evolution laws for the state variable have been investigated in a purely empirical manner because the microscopic definition of the state variable has not been known.
But we are now set out to derive evolution laws starting from the microscopic definition for the state variable defined by Eq. (\ref{theta}).

\subsection{Statistical properties}\label{thetaWPM}
As the state variable $\theta$ is defined in terms of the 0th weighted power mean, Eq. (\ref{theta}), the above discussions for the length constant $L$ may also apply to the state variable if the distribution function of $\theta_i$ is known.
For steady states, one may presume that the distribution function of $\theta_i$ is proportional to that of asperity-size distribution as $\theta_i=L_i/V$.
Following the same line of discussion in the previous subsection, one can conclude that the 0th weighted power mean is well approximated by the weighted mean if $2.5< \alpha <4$.

\subsection{General aspects}
\begin{figure}
\label{fig:sets}
\begin{center}
\includegraphics[scale=.5]{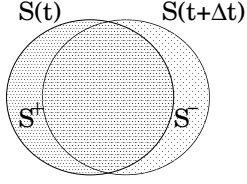}
\caption{The Venn diagram of asperity sets: Each circle represents ${\cal S}(t)$ (left) and ${\cal S}(t+\Delta t)$ (right), respectively.}
\end{center}
\end{figure}

Noting that the set of asperities is time-dependent due to sliding, we can write $\Delta \theta \equiv\theta(t+\Delta t) -\theta(t)$ as
\begin{equation}
\label{Dtheta}
\Delta\theta=\prod_{{\cal S}(t+\Delta t)}(\theta_i+\Delta t)^{\xi_i}-\prod_{{\cal S}(t)}\theta_i^{\xi_i}.
\end{equation}
Next we define ${\cal S}^{+} = {\cal S} (t+\Delta t) \setminus {\cal S} (t)$, which is the set of asperities that appear in the time window $[t, t+\Delta t]$.
In a similar manner, we also define ${\cal S}^{-} = {\cal S} (t) \setminus {\cal S} (t+\Delta t)$, which is the set of asperities that disappear in $[t, t+\Delta t]$.
These sets are illustrated in Fig. \ref{fig:sets}.

For simplicity, we assume that $\sum_{\cal S^{-}} A_i(0) = \sum_{\cal S^{+}} A_i(0)$ so that $\sum_{\cal S} A_i(0)$ is time-independent.
We further assume that $\sum_{\cal S^{-}} A_i(0)$ is proportional to $\sum_{\cal S} A_i(0)$ and to $V\Delta t/L$, where $L/V$ is interpreted as the average lifetime of asperities.
These assumptions are eventually expressed as
\begin{equation}
\label{renewrate}
\sum_{\cal S^{-}} A_i(0) = \sum_{\cal S^{+}} A_i(0)\simeq \frac{V\Delta t}{L} \sum_{\cal S} A_i(0).
\end{equation}

We then define a new weight coefficient $\nu_i$ that is normalized in ${\cal S}^{+}$ and 
${\cal S}^{-}$, since $\xi_i$ is no longer normalized in these new sets.
Using the assumptions made by Eq. (\ref{renewrate}), Eq. (\ref{xi}) leads to
\begin{eqnarray}
\label{nu1}
\xi_i &=& \frac{V\Delta t}{L} \nu_i \\
\label{nu2}
\nu_i &=& \frac{A_i(0)}{\sum_{\cal S^+} A_i(0)} = \frac{A_i(0)}{\sum_{\cal S^-} A_i(0)}.
\end{eqnarray}
Obviously, $\sum_{\cal S^+}\nu_i = \sum_{\cal S^-}\nu_i = 1$.

\subsection{Aging Law}
Then the aging law, Eq. (\ref{dieterich}), is obtained if the state variable $\theta$ can be approximated by the weighted mean of $\theta_i$.
\begin{equation}
\label{theta2}
\theta = \prod_{{\cal S}(t)} \theta_i^{\xi_i} \simeq \sum_{{\cal S}(t)} {\xi_i} \theta_i.
\end{equation}
The validity of this approximation is discussed quantitatively in section \ref{thetaWPM}.
Inserting Eqs. (\ref{nu1}) and (\ref{theta2}) into Eq. (\ref{Dtheta}), after a few lines of algebra one obtains
\begin{equation}
\label{evl1}
\Delta\theta(t) = \sum_{\cal S}\xi_i \Delta t  
+ \frac{V\Delta t}{L}\left(\sum_{\cal S^+} \nu_i \theta_i - \sum_{\cal S^-} \nu_i \theta_i \right).
\end{equation}
It is important to note here that the contact duration of asperity $\theta_i$ is generally much longer in ${\cal S}_-$ than in ${\cal S}_+$, as the asperities in ${\cal S}_+$ are "newborns" by definition.
This means that the second term on the right hand side of Eq. (\ref{evl1}) does not vanish.
As the asperities in ${\cal S^+}$ have just appeared in the time window $[t, t+\Delta t]$, their contact duration must be smaller than $\Delta t$.
Here we assume that
\begin{equation}
\label{ti_for_S+}
\theta_i\simeq\Delta t. \hspace{7mm} ({\rm for}\ \ i\in{\cal S^+})
\end{equation}
In addition, if the state of ${\cal S}^-$ may be approximated by that of ${\cal S}$,
\begin{equation}
\label{selfaverage}
\sum_{\cal S^-} \nu_i \theta_i \simeq\sum_{\cal S} \xi_i \theta_i= \theta,
\end{equation} 
Inserting Eqs. (\ref{ti_for_S+}) and (\ref{selfaverage}) into Eq. (\ref{evl1}), one obtains the aging evolution law.

\subsection{Slip Law}
The derivation of the slip law is not as straightforward as the aging law.
This is because the slip law does not explicitly incorporate the aging effect:
As apparent in Eq. (\ref{ruina}), the state variable $\theta$ can change only if $V\neq0$.
This contradicts the assumption that an asperity is strengthened with time.
Therefore, one must change the interpretation of $\theta_i$ in the logarithmic dependence of Eq. (\ref{loghealing}).

Here we reinterpret $\theta_i$ as $L_i/V(t_i)$, where $t_i$ is the time of the creation of asperity $i$.
In other words, the strength of asperity $i$ is determined by the slip velocity at the instance of creation, $V(t_i)$, and it is time-independent.
For instance, slowly-formed asperities are stronger, and the strength is unchanged during sliding.
This may be verified if the aging occurs only at the very early stage of asperity formation.
At this point we do not have any verification, but admit this anyway to derive the slip law.

Because $\theta_i$ is time-independent, Eq. (\ref{Dtheta}) can be rewritten as
\begin{equation}
\label{evl2}
\Delta\theta=\prod_{{\cal S}(t+\Delta t)}\theta_i^{\xi_i}-\prod_{{\cal S}(t)}\theta_i^{\xi_i}.
\end{equation}
Using the notation of the asperity sets ${\cal S^+}$ and ${\cal S^-}$, one can write 
\begin{equation}
\label{evl3}
\prod_{{\cal S}(t+\Delta t)} \theta_i ^{\xi_i} - \prod_{{\cal S}(t)} \theta_i ^{\xi_i}
= \theta \left(\frac{\prod_{{\cal S}^+} \theta_i^{\xi_i}}{\prod_{{\cal S}^-} \theta_i^{\xi_i}} - 1\right).
\end{equation}
Inserting Eqs. (\ref{nu1}) and (\ref{evl3}) into Eq. (\ref{evl2}), one has
\begin{equation}
\label{evl4}
\frac{\Delta\theta}{\theta} 
= \left(\frac{\prod_{{\cal S}^+} \theta_i^{\nu_i}}{\prod_{{\cal S}^-} \theta_i^{\nu_i}}\right)
^{\frac{V\Delta t}{L}} - 1.
\end{equation}
As the microscopic state variable $\theta_i$ is determined by the slip velocity at the instance of the creation of asperity $i$, $\theta_i$ in ${\cal S^+}$ is given by $L_i/V$.
On the other hand, $\theta_i$ in ${\cal S^-}$ depends on the history of the slip velocity.
We assume that $\theta_i$ in ${\cal S^-}$ is approximated by $\theta$ on average:
$\prod_{{\cal S}^-}\theta_i ^ {\nu_i} \simeq\theta$.
These two assumptions lead to
\begin{equation}
\label{evl5}
\frac{\prod_{{\cal S}^+} \theta_i^{\nu_i}}{\prod_{{\cal S}^-} \theta_i^{\nu_i}} \simeq \frac{L}{V\theta}.
\end{equation}
Inserting Eq. (\ref{evl5}) into Eq. (\ref{evl4}), one obtains the slip law.
\begin{equation}
\label{evl6}
{\dot \theta} \simeq \frac{V\theta}{L}\log\frac{L}{V\theta}.
\end{equation}

\section{Values of friction coefficient and atomistic parameters} \label{section:alternative}
\subsection{Reference-state free formulation}
The RSF law as described by Eq. (\ref{rsf}) involves only the minute change in the friction coefficient with respect to a reference state.
As we shall show below, such a reference state is not essential indeed.
Within the framework described in section \ref{section:derivation}, one can construct an alternative form of the RSF law without any reference to steady states.
One can then discuss the absolute value of the friction coefficient.

Inserting Eqs. (\ref{activation}) and (\ref{Z}) into Eq. (\ref{fZ}), one has
\begin{equation}
\label{F_alt}
F = \left(\frac{E}{l}+\frac{k_BT}{l}\log\frac{V}{V_0}\right)\left(1+c\log\frac{\theta}{\tau}\right) \sum_{i\in {\cal S}} Z_i(0).
\end{equation}
Inserting Eq. (\ref{n_0}) into the above equation and divide it by the normal load $N$, one obtains an alternative expression for the friction coefficient that does not have any reference state.
\begin{equation}
\label{alt_law}
\mu(V, \theta) = \frac{E}{P\Omega}\left(1+\frac{k_BT}{E}\log\frac{V}{V_0}\right) \left(1+c\log\frac{\theta}{\tau}\right),
\end{equation}
where we use Eqs. (\ref{P}), (\ref{Omega}), and (\ref{a_simple}).
Importantly, the definition of the state variable $\theta$ here is also given by Eq. (\ref{theta}) and therefore the time evolution of $\theta$ must be common to the conventional case.

The steady-state behavior of Eq.  (\ref{alt_law}) is obtained by setting $\theta=L/V$.
\begin{equation}
\label{alt_ss}
\mu_{\rm ss}(V) = \frac{E}{P\Omega}\left(1+\frac{k_BT}{E}\log\frac{V}{V_0}\right) \left(1+c\log\frac{L}{V\tau}\right).
\end{equation}
By differentiating the above equation with respect to $\log V$, one can retain Eq. (\ref{a-b}).
As $\tau$ is the shortest time scale needed for frictional aging, $L/\tau$ is regarded as the upper limit velocity for the RSF law to be valid.
Therefore, $\mu_{\rm ss}(L/\tau)$ approximates a typical experimental value of friction coefficient.
\begin{equation}
\label{typical_value}
\mu_{\rm ss}\left(\frac{L}{\tau}\right) = \frac{E}{P\Omega}
\left(1+\frac{k_BT}{E}\log\frac{L}{V_0\tau}\right).
\end{equation}

The friction law in the form of Eqs. (\ref{alt_law}) or (\ref{alt_ss}) does not involve any reference state, and the absolute value of the friction coefficient is expressed by the material constants, e.g., Eq. (\ref{typical_value}).
This makes a quite contrast to the conventional RSF law, which does not state anything about the absolute value of the friction coefficient.

If one further assumes that $c\ll 1$ and $E/k_BT \ge 200$, Eq. (\ref{alt_ss}) could be approximated as
\begin{equation}
\label{alt_ss_prev}
\mu_{\rm ss}(V) \simeq \frac{E}{P\Omega}+\frac{k_BT}{P\Omega}\log\frac{V}{V_0}
- \frac{cE}{P\Omega}\log\frac{V\tau}{L}.
\end{equation}
This is identical to the friction law obtained by Bar-Sinai et al. \cite{Bar-Sinai2014}, but it should be noted that one must assume large activation energy to obtain Eq. (\ref{alt_ss_prev}).

\subsection{Estimate of activation energy and activation volume}\label{activation_quantitative}
Equation (\ref{typical_value}) reveals that two nondimensional parameters that are defined from the three energy constants determine the absolute value of friction coefficient: $E/P\Omega$ and $E/k_BT$.
The real normal stress on asperities, $P$, may be the yield stress of protrusions.
One may assume that $P\simeq8$ GPa for most minerals \cite{Bowden}.
Then, quantitative estimate of the activation energy $E$ and the activation volume $\Omega$ is crucial for the friction coefficient.
The estimate requires an in-depth consideration on the atomistic processes at asperities.
Here we limit ourselves to semi-quantitative estimate with the aid of some experimental data.

First, noting that the second term in the bracket of Eq. (\ref{typical_value}) has only a minor contribution to the friction coefficient, $E/P\Omega$ must be close to $1$ for the friction coefficient to be a reasonable value:
\begin{equation}
\label{EPOmega}
E\simeq P\Omega.
\end{equation}
Then Eqs. (\ref{inequality_general}) and (\ref{EPOmega}) lead to
\begin{equation}
\label{kTPOmega}
k_BT \ll P\Omega.
\end{equation}
Recalling that $a\simeq k_BT/P\Omega$, this condition is equivalent to $a\ll 1$, which is consistent with typical values obtained in experiments \cite{Marone1998}.
Equation (\ref{EPOmega}) also means that $E/\Omega$ should be approximately $8$ GPa.

\begin{figure}
\label{fig:Emu}
\begin{center}
\includegraphics[scale=0.43]{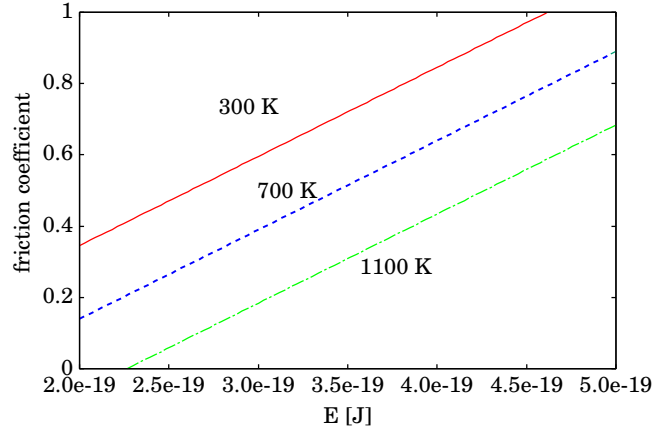}
\caption{Typical values of friction coefficient given by Eq. (\ref{typical_value}) are plotted as functions of the activation energy.
The red solid line, the blue dashed line, and the green dash-dotted line correspond to $T=300$, $700$, and $1100$ [K], respectively.
Here we assume that $P=8$ GPa and $\Omega =5.0\times10^{-29}$ ${\rm m}^3$, whereas the other parameters are chosen from Table I.}
\end{center}
\end{figure}

One can give a constraint on the activation energy from absolute value of friction coefficient.
Figure \ref{fig:Emu} shows the activation-energy dependence of the friction coefficient as described by Eq. (\ref{typical_value}).
At $T=300$ K, the friction coefficient takes plausible values ($0.6$ to $0.8$) only if the activation energy ranges from $3.0\times10^{-19}$ to
$3.8\times 10^{-19}$ J ($180$ to $230$ kJ/mol).
Despite some uncertainties in $P$ and $\Omega$,  one can see that possible values of the activation energy is actually limited to a very narrow range.

In rock friction experiments, the activation energy has been estimated as $180$ kJ/mol by assuming a relation similar to Eq. (\ref{alt_ss_prev}) \cite{Rice2001,Nakatani2001}.
This is consistent with the above estimate.
In another respect, the activation energy may be approximated by that for grain-boundary diffusion of silicon or oxygen atoms because the diffusion is realized through the reconnection of Si-O bonds in grain boundaries.
This activation energy has been experimentally estimated to be close to the above value \cite{Farver2000}.
Taking these observations into account, it would not be so bold to assume $E\simeq 200$ kJ/mol.
With this estimate, $E/k_BT$ is larger than $20$ for $T < 1200$ K (below $60$ \% the melting temperature), and therefore the assumption of thermal activation, Eq. (\ref{activation}), is justifiable in a wide range of temperatures.

Adopting $E\simeq 200$ kJ/mol and $P\simeq8$ GPa, it follows from Eq. (\ref{EPOmega}) that $\Omega \simeq 4 \times 10^{-29}$ m$^3$.
Note that this value is close to $5\times10^{-29}$ m$^3$ that was obtained via the temperature dependence of the parameter $a$ \cite{Nakatani2001}.

The activation volume $\Omega$ may be also estimated independently by an atomistic consideration as follows.
It is defined by Eq. (\ref{Omega}) as the product of the two quantities: the areal density of covalent bonds $n_0$ and a length constant $l$.
The latter may be the length of a covalent bond between Si and O and therefore estimated as $1.6 \AA$.
The former may be approximated by the number of covalent bonds contained in the atomistically narrow shear-zone that has the unit area and the thickness of $1.6 \AA$.
This gives $n_0 \sim 7\times 10^{18}$ m$^{-2}$, where the mass density of amorphous silica is assumed to be $2.2$ g/cm$^3$.
Then we obtain $\Omega\sim 2.3\times10^{-29} {\rm m}^3$.
Though crude estimation, this is comparable to the above experimental values.

\section{Concluding Remarks}\label{section:conclusions}
\subsection{Summary of parameters}
We have derived the RSF law based on the microscopic constitutive laws for asperities.
The essential physical processes at asperities are covalent-bond reconnection by the shear 
and time-dependent increase of the covalent bonds.
As a consequence, expressions for $a$ and $b$ are given in terms of atomistic parameters such as activation energy and activation volume; and material constants such as yield stresses and the sound velocity.
See Eqs. (\ref{a}) and (\ref{b}).
We retain the well-known expression for $a$ as far as the extent of aging is small: $c\ll 1$.
The microscopic expression for $b$ is indeed more complicated and can depend on the arbitrary reference velocity, although the dependence is insignificant as far as $c\ll1$.
Therefore, $c\ll 1$ is a vital condition for the RSF law.

As the covalent-bond reconnection is a thermal activation process, the activation energy $E$ plays an essential role in the RSF law.
The activation energy must be much larger than $k_BT$ not only because of the definition of thermal activation:
This is also required for the RSF law to be valid.
For instance, if $E/k_BT= 5$, parameter $b$ could not be constant as apparent from Eq. (\ref{b}). 
Here we estimate the activation energy as $200$ kJ/mol.
Then $E/k_BT\simeq 81$ at $T=300$ K; and it is larger than $20$ unless the temperature is higher than $1200$ K.

To derive the RSF law we assume some additional inequalities: Eqs. (\ref{inequality_general}) and (\ref{kTPOmega}).
As Eq. (\ref{sigma_i}) leads to ${\bar f}l \simeq \sigma_i\Omega$, Eqs. (\ref{inequality_general}) and (\ref{kTPOmega}) can  be combined into
\begin{equation}
\label{inequality_summary}
k_BT < \sigma_i\Omega < E\simeq P\Omega.
\end{equation}

\subsection{Scale Dependence of the Length Constant $L$}
We also derive the expression for the length constant $L$ and clarify the statistical nature.
As $L$ scales the critical nucleation size of unstable rupture, it is one of the most important quantities in the literature of earthquake physics.
We establish the precise definition of $L$ in the form of Eq. (\ref{L}): it is defined as the 0th weighted power mean of a linear dimension of asperities.
The scaling properties of $L$ is significantly different from the arithmetic mean and the weighted mean as discussed in section \ref{section:WPM}.
The difference between the 0th weighted power mean and other means depends on the statistical nature of the asperity size, and is enhanced if the size distribution is a power law.
If the exponent for the size distribution is smaller than $3$, the length constant $L$ is scaled with the upper cutoff length.
The apparently large critical slip distance observed in natural faults may be explained in terms of this scaling property, although the estimate of the exponent may be difficult for natural faults.
Such a topographical study of natural faults would be also a subject of the future work.

At least, in experiments, one can estimate the exponent $\alpha$ from a direct observation of asperities.
\cite{Dieterich1996} reported $\alpha$ ranging from $2.2$ to $4.2$.
(Note the relation $\alpha=2D-1$, where $D$ is the exponent for the area distribution.)
The constant $L$ may differ significantly within this range and therefore our result may be tested directly in such an observations together with the measurement of friction.
Such an experimental verification of the scaling properties of $L$ would be an intriguing problem to be investigated in the future.

\subsection{Evolution Laws}
The mathematical expression for the state variable is given here.
It provides a starting point for the systematic derivation of evolution laws.
As a demonstration, we have derived the two evolution laws: both the aging and the slip laws.
Derivation of some other known evolution laws would be an interesting problem to be investigated. 
Furthermore, starting from Eq. (\ref{theta_simple}), one can try to derive a novel evolution law theoretically.
Such attempts will be published elsewhere.

\begin{acknowledgments}
The author acknowledges helpful comments by Miki Takahashi, Takehiko Hiraga, Naoyuki Kato, Hiroshi Matsukawa, Mark O. Robbins, and Tetsuo Yamaguchi.
He also acknowledges JSPS Grand-in-Aid for Scientific Research.
\end{acknowledgments}

\label{lastpage}
\end{document}